\documentclass[runningheads]{svmult}
\usepackage{makeidx}   % allows index generation
\usepackage{graphicx}  % standard LaTeX graphics tool
\usepackage{subeqnar}  % subnumbers individual equations
\usepackage{multicol}  % used for the two-column index
\usepackage{cropmark} % cropmarks for pages without
\usepackage{physprbb}  % centered layout of diverse elements, etc.
\makeindex             % used for the subject index
\begin{document}
\title*{Search for Physics Beyond Standard Model\protect\newline
at HERA}
\toctitle{Search for Physics Beyond Standard Model at HERA}
% allows explicit linebreak for the table of content
%
%
\titlerunning{Search for Physics Beyond Standard Model at HERA}
% allows abbreviation of title, if the full title is too long
% to fit in the running head
%
\author{Masahiro Kuze \, \,({\it e-mail:} masahiro.kuze@kek.jp)}
\authorrunning{Masahiro Kuze}
% if there are more than two authors,
% please abbreviate author list for running head
%
%
\institute{Institute of Particle and Nuclear Studies, KEK,
Tanashi, 188-8501 Tokyo, Japan}

\maketitle              % typesets the title of the contribution

\begin{abstract}
The latest status\index{abstract} of searches at HERA for physics beyond
the Standard Model is summarized on behalf of the H1 and ZEUS collaborations.
Emphasis is put on production of resonant particles accessible within
the HERA center-of-mass energy, such as leptoquarks, squarks in
$R$-parity-violating supersymmetry or excited fermions.
%Both collaborations have accumulated a large statistics of positron-proton
%collisions, on which the majority of the results are based.
Most of the results presented here are based on the full available statistics
of positron-proton collisions, and
also preliminary results from very recent electron-proton running are
presented.
%The contents include some updated results made available since the time of
%the workshop.
Results which have been updated since this Ringberg '99 Workshop
are also included.
\end{abstract}

\section{Introduction}
The HERA collider at DESY is a unique facility where electrons (or positrons)
and protons interact at a high center-of-mass energy $(\sqrt{s})$, thus
allowing to probe very short distances.
An electron or positron beam of 27.5\,GeV collides with a proton beam of
820\,GeV (920\,GeV since 1998), resulting in $\sqrt{s}=300$\,GeV (318\,GeV).
New phenomena in lepton-quark interactions
due to physics at large energy scale could be observed at this
unpreceded energy.
Unlike $e^+e^-$ or $\bar{p}p$ colliders, the initial state at HERA
has both non-zero lepton and baryon numbers, which makes searches
at HERA most powerful in discovering particles which carry
both of these quantum numbers.

The H1 and ZEUS experiments have been taking data at HERA since 1992.
Most of the results presented here are based on the
large data sample of $e^+p$ collisions taken between 1994--1997,
corresponding to luminosities of
36.5\,pb$^{-1}$ and 47.7\,pb$^{-1}$ for H1 and ZEUS,
respectively.  From 1998 to spring 1999, HERA provided $e^-p$
collisions with the increased proton beam energy.  Preliminary
results from 16\,pb$^{-1}$ of data taken in this period are also
presented.
 
\begin{figure}[tb]
\begin{center}
\includegraphics[width=.85\textwidth]{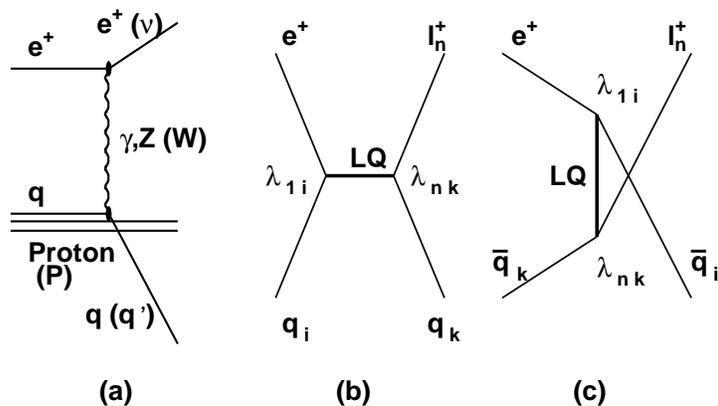}
\end{center}
\caption{Diagrams of (\textbf{a}) deep inelastic scattering of $e^+p$;
(\textbf{b}) s-channel and (\textbf{c})u-channel LQ processes involving
$F=0$ LQs}
\label{kuze:lqdiag}
\end{figure}

For most of the searches, major Standard Model (SM)
background comes from neutral-current (NC)
and charged-current (CC) deep inelastic scattering (DIS) processes, as
illustrated in Fig.~\ref{kuze:lqdiag} (a).
The NC process occurs via t-channel exchange of $\gamma$ or $Z$ boson, while
the CC process turns the electron into a neutrino through $W$-boson exchange.
The following kinematic variables are frequently used in DIS analyses:

\begin{eqnarray}
Q^2 = -q^2 = -(k-k')^2 \;,\\
x = Q^2/(2q\cdot p) \;, \\
y = (q\cdot p)/(k \cdot p) \;, 
\end{eqnarray}
where $k$ and $k'$ are the four-momenta of the incoming and outgoing lepton,
respectively, and $p$ is the four-momentum of the incoming proton.
$Q^2$ is the negative square of the momentum transfer $(q)$, $x$ is
the Bjorken scaling 
variable and $y$ is sometimes called the inelasticity parameter.

\section{Leptoquarks and Related Searches}
Leptoquarks (LQs) have both lepton $(L)$ and baryon $(B)$ numbers and couple
directly to lepton-quark pairs.
They appear in many models beyond SM, such as
Grand Unified Theories, and naturally relate the two families of
fermions: leptons and quarks.
They carry color charges like quarks and have fractional electric charges.
The set of LQs which preserves $SU(3)\times SU(2) \times U(1)$
symmetry has been specified by Buchm\"uller, R\"uckl and Wyler
(BRW)~\cite{kuze:BRW}.
They can be classified into scalar and vector LQs, and into
$F=0$ and $|F|=2$ LQs, where $F=L+3B$ denotes the fermion number.
\subsection{LQ Production and Decay at HERA}
Figures \ref{kuze:lqdiag} (b) and (c) show LQ processes at HERA with
s- and u-channel exchange of a LQ,
with possible lepton- and/or quark-generation-changing interactions.
%The diagrams are generalized to allow for generation-changing LQs
%(the final-state lepton and/or quark can be of a different generation
%from the initial state), which will be discussed in Sect.~\ref{kuze:LFV}.
Lepton-flavor-violating (LFV) LQs will be discussed in Sect.~\ref{kuze:LFV},
and only first-generation LQs ($n=1$ and $i=k=1$ for the Yukawa couplings
$\lambda$ in the diagrams) are discussed here.
According to the BRW model, the decay branching ratio of LQ to
$eq$ is either 100\% or 50\%, depending on the LQ quantum numbers.
In the latter case, the remaining decay is to a $\nu q$ final state.
At HERA, highest sensitivity to first-generation LQs can be achieved
by producing the LQs via a fusion between the lepton and the valence quarks
$(u,d)$, i.e. for $F=0$ $(|F|=2)$ LQs in $e^+p$ $(e^-p)$ collisions.

An individual LQ event has exactly the same topology as a NC or CC DIS event.
If the LQ mass, $m_{\mathrm LQ}$, is smaller than $\sqrt{s}$ and 
the coupling $\lambda$
is not too large (order of unity or less), the s-channel resonant production
dominates.
%\footnote{otherwise the u-channel diagram and interference terms
%between the SM and LQ diagrams become non-negligible; see
%e.g.~\cite{kuze:matsushita}}.
In this case there appears a sharp peak at $m_{\mathrm LQ}$
in the $eq$ or $\nu q$ invariant
mass distributions, or at $x_0=m_{\mathrm LQ}^2/s$
in the DIS variable $x$.
The production cross section for this case can be simply approximated with
the narrow-width approximation (NWA):
\begin{equation}
\sigma(ep \to LQ\; X) = (J+1)\frac{\pi}{4s}\lambda^2 q(x_0, m_{\mathrm LQ}^2)\;,
\end{equation}
where $\lambda$ is the Yukawa coupling, J is the spin of the LQ and $q$
is the quark distribution function evaluated at the resonance $x$=$x_0$
with the scale $Q^2 = m_{\mathrm LQ}^2$.
Another characteristic of LQ process is the different $y$ distribution
as compared to DIS.
The s-channel production of a scalar LQ has a flat $y$ distribution, while
it is $(1-y)^2$ for the vector case.
They are contrasted to the $1/y^2$ dependence of DIS processes at fixed $x$.
Therefore, the search strategy is to start from a selection of NC or CC DIS
events and then to look for a resonance peak in the large-$y$ region.

The signal of a LQ with mass close to or above $\sqrt{s}$ looks different
from what has been discussed above,
but the sensitivity on such LQs does not vanish~\cite{kuze:matsushita}.
This will be discussed in Sect.~\ref{kuze:CI}.

\subsection{Mass Distribution and Limits}
\begin{figure}[tb]
\begin{center}
\includegraphics[width=\textwidth]{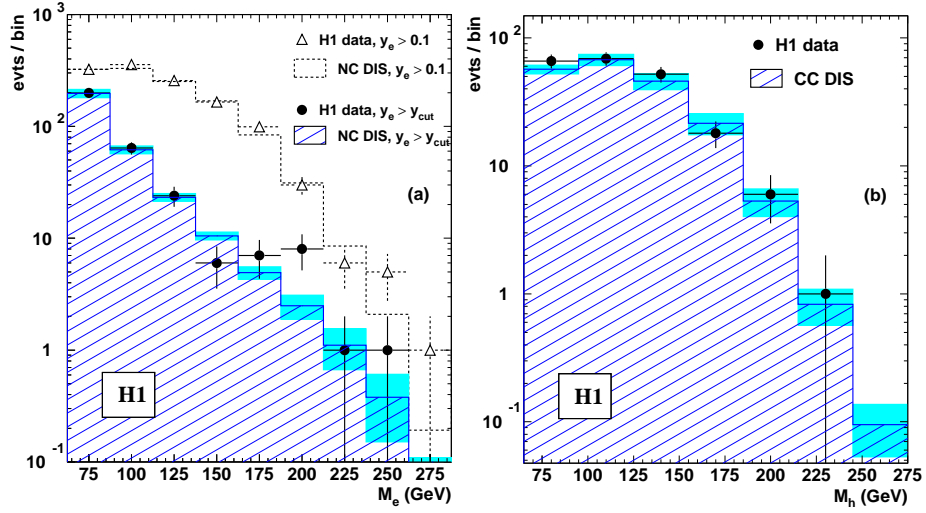}
\end{center}
\caption{Mass spectum for (\textbf{a}) NC-like and 
(\textbf{b}) CC-like event selection from H1 $e^+p$ data ({\it points\,})
compared with DIS expectations ({\it histograms\,}).
In (\textbf{a}), the distribution is shown before and after the mass-dependent
lower cut in $y$ designed to enhance the signal significance.
The shaded boxes on the histograms indicate the uncertainties on
the expectations}
\label{kuze:h1lqmass}
\end{figure}

Figure~\ref{kuze:h1lqmass} shows the mass distribution from the H1
analysis~\cite{kuze:h1lq} of 1994--1997 $e^+p$ data.
For the NC DIS selection, the mass is
calculated using the energy and angle of the scattered positron,
and for the CC DIS selection using the hadronic variables with the
Jacquet-Blondel method~\cite{kuze:JB}.
Slight excess of events is observed in the NC channel
at large $y$ around 200\,GeV, which
comes mainly from the data taken during 1994--1996 and was particularly
noteworthy before the 1997 data were added~\cite{kuze:h1excess}.
No significant excess was seen in the 1997 data alone, and overall
significance has decreased in the whole data sample.
The mass distribution in the CC channel is in good agreement with DIS
expectation within the uncertainty.

\begin{figure}[tb]
\begin{center}
\includegraphics[width=\textwidth]{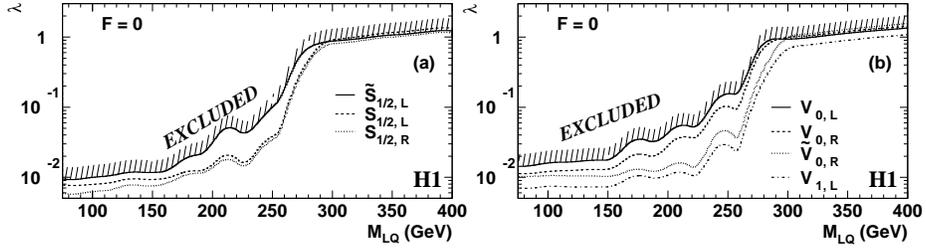}
\end{center}
\caption{Upper limits at 95\% CL on Yukawa couplings $\lambda$ as a function of
the LQ masses for (\textbf{a}) scalar and (\textbf{b}) vector $F=0$ leptoquarks.
In the plots shown here, only NC data have been used}

\label{kuze:h1lqlim}
\end{figure}

From the observed and expected mass spectrum, the limits at 95\%
confidence level (CL) on the
Yukawa coupling $\lambda$ can be obtained for each LQ type,
as shown in Fig.~\ref{kuze:h1lqlim}.
At an electromagnetic strength $\lambda \sim 0.3$, the limits on
scalar (vector) LQs extend up to 275\,GeV (284\,GeV).
Here only limits on $F=0$ LQs are shown, but also results on $|F|=2$
LQs are obtained, with weaker limits than $F=0$.

Also ZEUS has analysed $e^+p$ data in NC~\cite{kuze:zeuslq} and
CC~\cite{kuze:zeusvj} channels and obtained preliminary limits on LQs.
In the NC channel, ZEUS observes a few outstanding events
at large mass and large $y$, coming mainly from 1994--96
data~\cite{kuze:zexcess}, but the significance has decreased after adding
the 1997 data which more than doubled the total integrated luminosity.
More $e^+p$ data will be needed to clarify the origin of these high-mass, 
high-$y$ events events observed by H1 and ZEUS.

\begin{figure}[tb]
\begin{center}
\includegraphics[width=.6\textwidth]{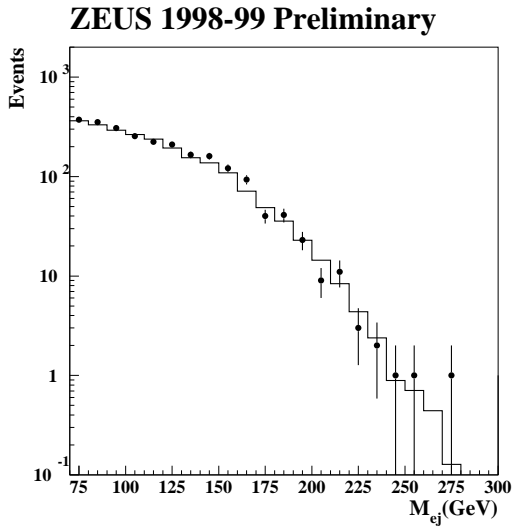}
\end{center}
\caption{Electron-jet invariant mass distribution from ZEUS $e^-p$ data
({\it points\,}) compared to NC DIS expectation ({\it histogram\,})}
\label{kuze:zlqmass}
\end{figure}

\begin{figure}[tb]
\begin{center}
\includegraphics[width=.9\textwidth]{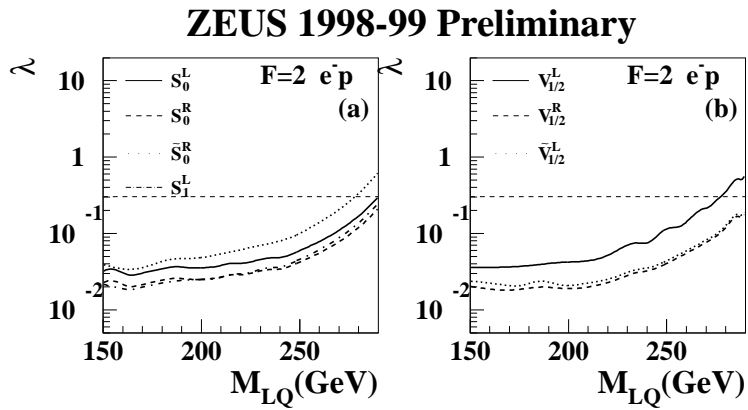}
\end{center}
\caption{Coupling limits at 95\% CL
as a function of the LQ masses for (\textbf{a}) scalar and
(\textbf{b}) vector $|F|=2$ leptoquarks.  The horizontal lines indicate the
coupling strength for an electroweak scale, $\lambda \sim 0.3$}
\label{kuze:zlqlim}
\end{figure}

ZEUS has also looked at the most recent $e^-p$ data taken during
1998--99~\cite{kuze:zejet}.
These data enhance very much the sensitivity to $|F|=2$
LQs, and the increased center-of-mass energy in addition extends
the sensitivity at the highest mass region ($\sim$ 300\,GeV)
close to the kinematical limit.
Figure~\ref{kuze:zlqmass} shows the mass distribution of events from NC DIS
selection.  Here the mass $(M_{ej})$ is the invariant mass between the
electron and jet, calculated directly using their energies and angles.
The agreement with SM DIS expectation is very good, and the limits on
Yukawa coupling for $|F|=2$ LQs are plotted in Fig.~\ref{kuze:zlqlim}.
At $\lambda \sim 0.3$, the exclusion limits extend up to about 290\,GeV
at 95\% CL.

\subsection{High-Mass Leptoquark and Contact Interactions}
\label{kuze:CI}
The LQ mass range in 
Fig.~\ref{kuze:h1lqlim} extends beyond the center-of-mass energy
300\,GeV up to 400\,GeV.  The H1 analysis takes into account not only
the s-channel contribution but also u-channel and interference terms
between the LQ and SM DIS diagrams.  When the LQ mass is high, these
terms become important and actually the LQ is not resonantly produced but
gives virtual effects on DIS-like final states~\cite{kuze:matsushita}.
In this case, the mass spectrum would not have a narrow peak but a broad
mass range would be affected, and also at smaller $y$ values.
The range in mass and cut in $y$ are thus changed accordingly.
%to look at larger region of the phase space for the search of LQs with high mass.
On the other hand, the ZEUS analysis uses NWA and the quoted limits stop
before the coupling becomes too large.

When the mass is even higher, the effect of LQ can be described as an
effective four-fermion interaction known as $eeqq$ contact interaction (CI).
CI is an effective theory for physics at high mass scales in their
approximation at the low-energy limit, and can represent not only heavy LQs
but also a variety of models beyond SM such as new heavy vector bosons
or composite models of fermions.
The effect is parameterized by the ratio of the coupling and the mass
scale, $g/\Lambda$, and usually the convention $g^2$=$4\pi$ is adopted.
Depending on the chiral structure of the interaction, many CI models can
be constructed and modify the $eq \to eq$ cross section differently.
The common effect is the increase of cross section at high $Q^2$,
with the interference effect at intermediate $Q^2$ which can be constructive
or destructive depending on the model.

ZEUS has analysed 1994--97 $e^+p$ data in terms of 30 scenarios of CI
and derived lower limits on $\Lambda$.  For details of the analysis, refer to
\cite{kuze:zeusci}.  The limits at 95\% CL range between 1.7\,TeV to 5\,TeV
for the scenarios considered.  It is worth to note that $eeqq$ CI can be
tested also at LEP2 $(e^+e^- \to q\bar{q})$ and at Tevatron
$(q\bar{q} \to e^+e^-)$.  Generally the limits from competing colliders are
comparable, and in some cases ZEUS limits are most stringent or the
only existing limits.

H1 has released preliminary results on CI from $e^+p$ data~\cite{kuze:h1ci}
and interprets the limits also in terms of LQ mass and coupling ratio,
$m_{\mathrm LQ}/\lambda$.  The limits, valid for $m_{\mathrm LQ} \gg \sqrt{s}$,
range between 202\,GeV and 952\,GeV for various type of LQs.

\subsection{Lepton-Flavor Violation}
\label{kuze:LFV}
The processes in Figs.~\ref{kuze:lqdiag} (b) and (c) can allow
for the case $n>1$, where the outgoing lepton is a muon or tau lepton,
leading to an explicit lepton-flavor violation forbidden in the SM.
The event signature is striking and can be searched for with
little background from SM processes.

The H1 analysis~\cite{kuze:h1lq} observes no candidate event consistent
with $eq \to \mu q'$ or $eq \to \tau q'$ process, with the expected
background from SM being $0.12 \pm 0.05$ and $0.77 \pm 0.30$ events,
respectively.
The limits are interpreted in terms of a coupling for low-mass
LFV LQs, and also in terms of indirect effects of high-mass LFV LQs
like the CI analysis.
In the latter case, the limits are given on the quantity
$\lambda_{1i}\lambda_{nk}/m_{\mathrm LQ}^2$.
In some cases where second- or third-generation quarks are involved,
and especially for the $e$--$\tau$ LFV case, H1 gives more stringent limits
than low-energy processes such as rare decays of $\tau$ or $B$.

ZEUS has recently given preliminary results from a LFV LQ search in the muon
channel using 1994--97 data.  No candidate is found where 0.3 events
are expected from the SM background.  Limits have been obtained for the
low-mass LFV LQs.  Under the assumption that the couplings to $eq$ and
$\mu q'$ have the same electroweak strength, 95\% CL lower limit for
the mass of LQs extends up to~285\,GeV~\cite{kuze:zlfv}.

\section{Supersymmetry with $R$-parity Violation}
Supersymmetry (SUSY) is one of the most promising extension
of the SM.  Extensive searches are being performed at the high-energy
colliders LEP2 and Tevatron, and HERA is not an exception.
HERA's potential for SUSY discovery is maximal in the case of the
$R$-parity-violating (RPV) extension of Minimal SUSY Standard Model (MSSM).

\subsection{Phenomenology}
$R$-parity is a multiplicative quantum number defined as
$R$=$(-1)^{L+3B+2S}$, where $S$ is the spin of the particle,
and is assumed to be conserved in MSSM.
$R$ takes the value 1 for SM particles and $-1$ for their SUSY partners.
The consequences of the $R$-parity conservation are that
SUSY particles are always produced in pairs, and that
the lightest SUSY particle (LSP) cannot decay.

%When $R$-parity is violated, the SUSY superpotential gets additional
%terms~\cite{kuze:Barger}
However, the general supersymmetric and gauge-invariant superpotential
contains additional terms which violate $R$-parity~\cite{kuze:Barger}
\begin{equation}
    \lambda_{ijk}  L_{i}L_{j}\bar{E}_{k} +
    \lambda_{ijk}' L_{i}Q_{j}\bar{D}_{k} +
    \lambda_{ijk}''\bar{U}_{i}\bar{D}_{j}\bar{D}_{k}.
\end{equation}
Here $L$ and $Q$ denote the left-handed lepton and quark doublet superfields;
$\overline{E}$, $\overline{D}$, and $\overline{U}$ are the right-handed
singlet superfields for charged leptons, down-type quarks
and up-type quarks, respectively.
The indices $i,j,k$ denote the generation.
For each term and generation combination, a Yukawa coupling
$\lambda (\lambda', \lambda'')$ is introduced in the model as
an additional parameter.

Of particular interest for HERA is the second term with $\lambda_{1jk}'$,
which makes single production of a squark $(\tilde q)$
possible through the electron-quark fusion.
This process is very much like the scalar leptoquark production
discussed earlier.  If the squark decays to $eq$ with the same
Yukawa coupling, the analysis is exactly the same as in the LQ search.
However, there are also $R$-parity-conserving decays with gauge
couplings $\tilde q \to q \chi^0$ and $\tilde q \to q' \chi^\pm$.
These decays to neutralino/chargino compete with the LQ-like decay,
and the branching ratio depends on the unknown Yukawa coupling
and on the MSSM parameters.

The neutralino or chargino decays subsequently to a lighter gaugino
(cascade decay), or directly through the RPV coupling.
Even the LSP, usually taken to be $\chi_1^0$, decays to
SM particles via $\chi_1^0 \to e^\pm qq$ and $\chi_1^0 \to \nu(\bar \nu) qq$.
Therefore, there are variety of final states of squarks involved in
the RPV SUSY phenomenology.

\subsection{Experimental Results}
The preliminary ZEUS analysis of 1994--97 $e^+p$ data~\cite{kuze:zrpv} makes
a simplifying
assumption that $\tilde q \to q \chi_1^0$ dominates the gauge decay
and ignores cascade decays and charginos.  The nature of $\chi_1^0$ is
assumed to be a pure photino, in which case the branching ratios to
$eq$ and $q\chi_1^0$ decays follow a simple formula~\cite{kuze:BD}.

The final states are classified to $eq$, $e^+qqq$ and $e^-qqq$.
The second topology looks like a NC event with multiple jets in the
hadronic final state.  The last topology has a ``wrong sign'' lepton
for the $e^+p$ collision and is a very clean channel with small background.
The $\nu (\bar \nu) qqq$ final states, contributing 12\% of the $\chi_1^0$ decay,
is not investigated in this analysis.

\begin{figure}[tb]
\begin{center}
\includegraphics[width=.7\textwidth]{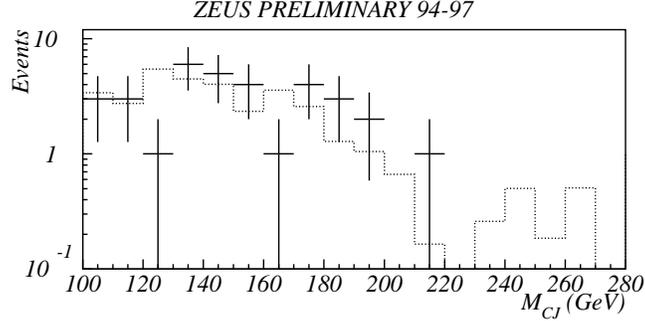}
\end{center}
\caption{Mass distribution of events selected in $e^+qqq$ channel for
data ({\it crosses\,}) and expectation ({\it dashed histogram\,})}
\label{kuze:zrpvmass}
\end{figure}

\begin{figure}[tb]
\begin{center}
\includegraphics[width=.7\textwidth]{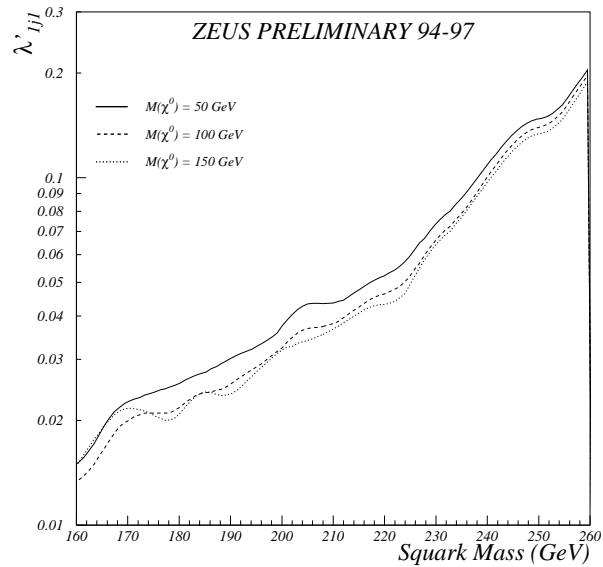}
\end{center}
\caption{The 95\% CL upper limits on the RPV Yukawa couplings $\lambda_{1j1}'$
as a function of the mass of the squark $\tilde u_j$ for different neutralino 
masses $M(\chi^0)$}
\label{kuze:zrpvlim}
\end{figure}

Figure~\ref{kuze:zrpvmass} shows the mass distribution of ZEUS events
passing the cuts for $e^+qqq$ channel.  Here, 33 events are observed while
33.6 events are expected from SM, mainly from NC DIS and a small contribution
from photoproduction processes.  In the $e^-qqq$ channel, no event is
observed while 0.06 events is expected from the background.
Since no signal for a resonance is found,
limits are derived on the couplings
(assuming only one $\lambda_{1j1}'$ to be non-zero at a time)
as a function
of the squark mass, shown in Fig.~\ref{kuze:zrpvlim}.  Here $j=1,2,3$
corresponds to $e^+d \to \tilde u, \tilde c, \tilde t$ production,
respectively.  The analysis is repeated for three different
$\chi_1^0$ masses.

\begin{figure}[tb]
\begin{center}
\includegraphics[width=.75\textwidth]{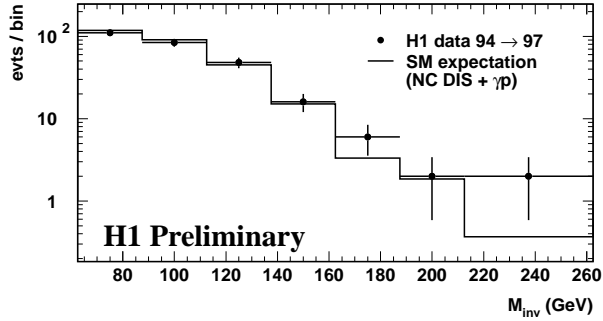}
\end{center}
\caption{Mass spectrum for $e\;+\;multijets$ final states for data
({\it points\,}) and SM expectation ({\it histogram\,})}
\label{kuze:h1rpvmass}
\end{figure}

\begin{figure}[tb]
\begin{center}
\includegraphics[width=.75\textwidth]{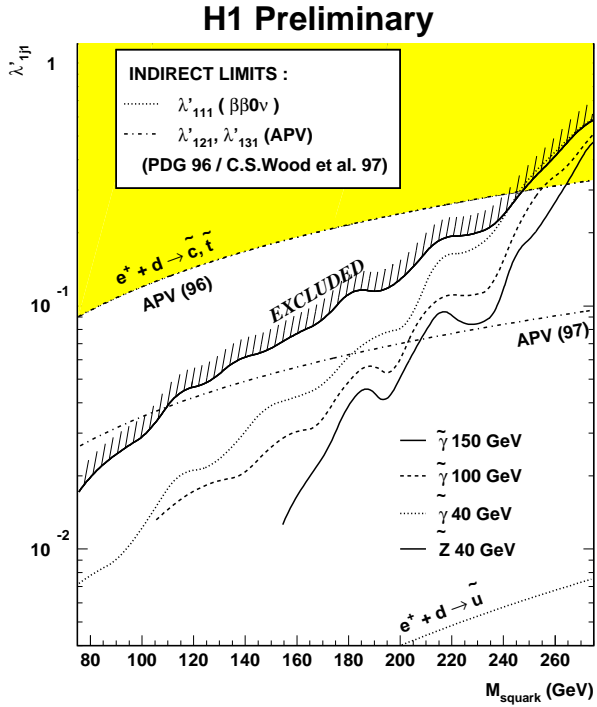}
\end{center}
\caption{Upper limits at 95\% CL for the RPV couplings $\lambda_{1j1}'$
as a function of the squark mass, compared with most stringent indirect limits}
\label{kuze:h1rpvlim}
\end{figure}

The preliminary H1 analysis of 1994--97 $e^+p$ data~\cite{kuze:h1rpv} takes
neutralino mixing into account and investigates SUSY parameter space
of a photino-dominated neutralino and a zino-dominated neutralino.
It also calculates the branching ratio of cascade decays with $\chi_2^0$ and
$\chi_1^+$.
Figure~\ref{kuze:h1rpvmass} shows the mass distribution in the $e\;+\;multijets$
channel.  With softer selection cuts than those used in the ZEUS analysis,
289 candidates are observed while the SM expectation is
$285.7 \pm 28.0$ events.
In the ``wrong sign'' channel, one candidate passes the selection
while the background is expected to be $0.49 \pm 0.2$.
No evidence for a squark is found, and
the limit on the Yukawa coupling is shown in Fig.~\ref{kuze:h1rpvlim}
for three different masses of photino-dominated neutralino and one case of
zino-dominated neutralino.  The coupling $\lambda_{111}'$ is strongly
constrained by the neutrino-less double-beta decay, but the results on 
$\lambda_{121}'$ and $\lambda_{131}'$ are competitive with
atomic-parity-violation experiments.

\section{Excited Fermions}
The composite models of fermions regard them as being built
from more fundamental particles.  In such models, leptons and/or quarks
can be excited to a higher-mass state and decay ``radiatively'' to the
stable ground state (normal fermions), emitting gauge bosons such
as photon, $W$ or $Z$ bosons.  At HERA, excited states of electrons
or quarks can be created through the $t$-channel photon or $Z$ exchange
in $eq$ interaction, and excited neutrinos or quarks can be created through the
$W$ exchange.  The excited electrons can also be produced in the elastic
process $ep \to e^*p$.
The search strategy is to reconstruct a photon, $W$ or $Z$
boson $(V)$ in an event and look for a resonance peak in the $f-V$ invariant
mass, where $f$ is an electron, a quark or a neutrino (missing momentum).

\begin{figure}[b]
\begin{center}
\includegraphics[width=.5\textwidth]{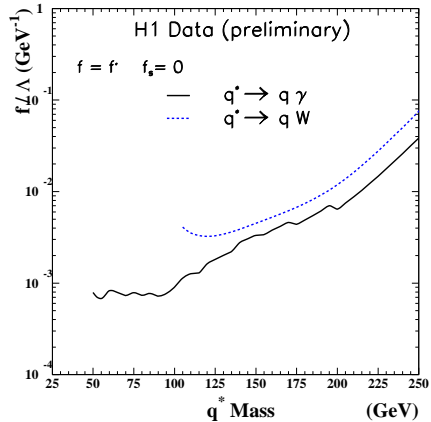}
\end{center}
\caption{Upper limit at 95\% CL for the excited quark coupling divided
by the compositeness scale, $f/\Lambda$, as a function of the $q^*$ mass}
\label{kuze:h1qstar}
\end{figure}

The preliminary H1 analysis of 1994--97 $e^+p$ data~\cite{kuze:h1ef} searches for
excited fermions in the decay channels $e^* \to e \gamma, e Z, \nu W;
\nu^* \to \nu \gamma, \nu Z, e W; q^* \to q \gamma, q' W$.  The hadronic decays
of $W, Z$ and leptonic decays $W \to e \nu; Z \to e \bar e, \nu \bar \nu$ have
been exploited.
In all cases, the numbers of observed events are in agreement with the
SM expectations and no evidence for a resonance has been found.

The derived limits are based on a specific phenomenological model
of excited fermions~\cite{kuze:eftheo} in which the cross sections depend
on coupling constants $f, f'$ and $f_s$ for the gauge groups $SU(2), U(1)$
and $SU(3)$, respectively, and the compositeness scale $\Lambda$.
The decay branching ratio of excited fermions are determined once
relationships between the couplings are fixed.
Figure~\ref{kuze:h1qstar} shows, as an example, the limit on $f/\Lambda$ for
the excited quark production under the assumption $f=f'$ and $f_s=0$.
The latter condition makes the results
complementary to the searches at Tevatron, where the production of $q^*$
is assumed to occur through the quark-gluon fusion
$(f_s \neq 0)$~\cite{kuze:cdf}.

\begin{figure}[tb]
\begin{center}
\includegraphics[width=.6\textwidth]{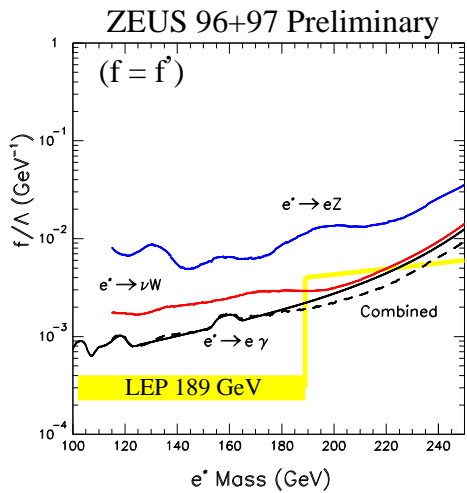}
\end{center}
\caption{Upper limits on $f/\Lambda$ at 95\% CL for $e^*$ 
production, as a function of the $e^*$ mass.
The combined limit from the three decay channels  ({\it dashed line\,})
is also shown}
\label{kuze:zestar}
\end{figure}

\begin{figure}[tb]
\begin{center}
\includegraphics[width=.6\textwidth]{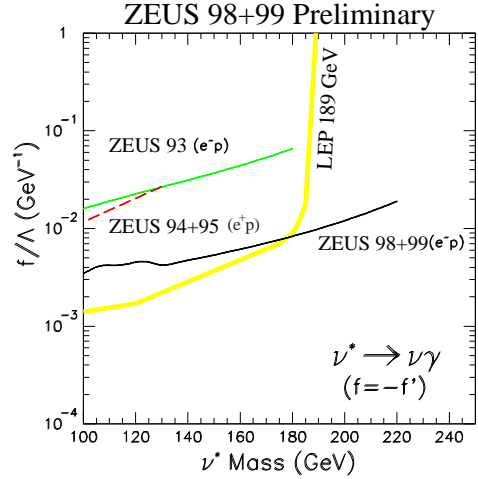}
\end{center}
\caption{Upper limits on $f/\Lambda$ at 95\% CL for $\nu^*$
production, as a function of the $\nu^*$ mass}
\label{kuze:znstar}
\end{figure}

ZEUS has made a preliminary search for $e^*$ using the data taken during
1996--97 (37\,pb$^{-1}$)~\cite{kuze:zef}.
The decays $e^* \to e \gamma, e Z, \nu W$
are exploited with the hadronic decay of $W$ and $Z$.
No evidence for a resonance has been found, and the limits on $f/\Lambda$
using the same model~\cite{kuze:eftheo} have been derived under the assumption
$f=f'$.  The limits are shown in Fig.~\ref{kuze:zestar}.  A combined limit
from the three decay channels is also derived.  It can be seen that limits
from LEP2 are more stringent below its center-of-mass energy, and HERA
limits are competitive above it, where the on-shell production of $e^*$ is not
possible at LEP2 and limits come from searching for the signs of 
virtual $e^*$ exchanges in
the process $e^+e^- \to \gamma \gamma$~\cite{kuze:delphi}.

ZEUS has also performed a preliminary search for $\nu^*$ using the recent
$e^-p$ data taken in 1998--99~\cite{kuze:zef}.
Using this smaller integrated luminosity (16\,pb$^{-1}$) than $e^+p$ data is far
more beneficial in $\nu^*$ search, since the production cross section
in $e^+p$ at high $\nu^*$ mass is strongly suppressed compared to $e^-p$.
It is due to the smaller $d$-quark density compared to $u$ quark at high
$x$ and the $(1-y)^2$ suppression factor coming from the chiral nature
of $W$ exchange.
The search has been done in $\nu^* \to \nu \gamma$ channel only, and 
two observed events are in agreement with the SM expectation of $1.8 \pm 0.2$.
The obtained limit on $f/\Lambda$ for the case $f = -f'$
(the $\nu^* \to \nu \gamma$ decay vanishes for the case $f = f'$) is shown in
Fig.~\ref{kuze:znstar}.

\section{Conclusion and Outlook}
Extensive searches for physics beyond the Standard Model are being performed
by the two collaborations H1 and ZEUS at HERA.
No convincing signal of new particles has been established so far, giving
limits on their production which are competitive with the searches
at other colliders.

HERA is planning to continue running until May 2000 with the current design
and then undergoes major upgrade plans in order to increase the luminosity.
There will be new final-focusing magnets close to the interaction point,
which means that also the detectors will be modified
during the upgrade shutdown.

The new running from 2001, with five times more luminosity than the current
design value, will bring an order of 1\,fb$^{-1}$ of integrated luminosity.
%within $\sim$~5~years.
This large amount of data allows the experiments to make high-statistics
analyses at large-$Q^2$ and high-mass regions, which will eventually
unreveal the breakdown of Standard Model if the new physics is existing
within the reach of HERA.

%INDEX%%%%%%%%%%%%%%%%%%%%%%%%%%%%%%%%%%%%%%%%%%%%%%%%%%%%%%%%%%%%%%%
% Please check with the editor of your book whether he plans to
% include a "mutual" subject index - if so, please code your entries
% in the standard syntax. For your own purposes you may print your
% "personal" index by using the following commands:
%
\clearpage
\addcontentsline{toc}{section}{Index}
\flushbottom
\printindex
%%%%%%%%%%%%%%%%%%%%%%%%%%%%%%%%%%%%%%%%%%%%%%%%%%%%%%%%%%%%%%%%%%%%%

\end{document}